# Fluctuation Pressure Assisted Ejection of DNA From Bacteriophage

Michael J. Harrison

*Department of Physics and Astronomy, Michigan State University*
*East Lansing, MI 48824-2320, USA*

**Abstract.** The role of thermal pressure fluctuation excited within tightly packaged DNA prior to ejection from protein capsid shells is discussed in a model calculation. At equilibrium before ejection we assume the DNA is folded many times into a bundle of parallel segments that forms an equilibrium conformation at minimum free energy, which presses tightly against internal capsid walls. Using a canonical ensemble at temperature T we calculate internal pressure fluctuations against a slowly moving or static capsid mantle for an elastic continuum model of the folded DNA bundle. It is found that fluctuating pressure on the capsid internal wall from thermal excitation of longitudinal acoustic vibrations in the bundle may have root-mean-square values which are several tens of atmospheres for typically small phage dimensions. Comparisons are given with measured data on three mutants of lambda phage with different base pair lengths and total genome ejection pressures.

## INTRODUCTION

It has been emphasized that thermal fluctuations away from equilibrium in small systems assume increasingly important roles as the system size becomes ever smaller and approaches molecular dimensions [1]. In the case of viral entities, where the phage size may very considerably [2,3], thermal pressure fluctuations may become exerted internally by confined DNA that is already tightly packged in an equilibrium conformation within a capsid mantle. Such pressure fluctuations can be expected to play a signigicant role in the initial packaging and subsequent dynamics of DNA ejection from bacteriophage [4,5,6,7,8] when the virus engages a host cell surface receptor and transfers its DNA into the cell.

Geometrical and topological constraints on the structure and size of viral capsids were discussed in an early paper by Caspar and Klug [9] almost fifty years ago. More recent work [2,10,11] has discussed highly symmetric capsid structure in the form of nanometer-sized protein shells. And viral infectivity has been related to DNA length and capsid size [12]. It has also been concluded that very large conformal changes may occur in some capsid shells when their viral genomes become tightly packaged [13]. We shall adopt a simpler model of viral capsids that contain packaged genomes than has generally been discussed [14].

## CALCULATION OF MODEL ENCAPSIDATED DNA AND CAPSID

Consider a circular cylindrical bundle of folded double-stranded DNA bacteriophage segments with cross-section diameter L and equal segment lengths L, packaged coaxially and situated symmetrically within a cylindrical capsid cavity of diameter (L+2h) and length L, where h is the thickness of the empty annular region surrounding the cylindrical DNA bundle  The capsid cavity is then a circular cylinder of diameter (L+2h) and length L. If we assume that the capsid cavity has twice the volume [15] of the DNA bundle, then geometric considerations  lead to the relation h=L / [2(1+√2)] = 0.2071/L, so that (L+2h) =1.414 L is the model capsid diameter. Capsids are nanometer sized in general magnitude [10], but vary greatly [2,11] depending specifically on the DNA genomes they carry within. We shall adopt representative capsid diameters near 50 nm [5,6,21] in our calculation of fluctuation pressures exerted between a vibrating DNA bundle and the capsid shell to which it is tethered.

Our discussion of the role of fluctuations rests on a picture of a tightly packaged encapsidated DNA bundle that acquires its thermal equilibrium structure through variational minimization of the free energy [7] of the capsid-DNA system. Fluctuations of the DNA bundle about its equilibrium conformation takes place in the form of thermally excited longitudinal acoustic vibrations. These vibratory  excitations in the DNA bundle will be regarded as taking place with the bundle bounded tightly by capsid walls which themselves continue to maintain positions close to their thermal equilibrium geometry, and do not significantly participate in thermal excitation on the same scale as the DNA segment bundles. The bundles are assumed to be connected fixedly at both ends to their capsid mantles.

In order to calculate the thermal fluctuations in pressure at the locations where DNA bundles are attached to their capsid shells we shall approximate a bundle of folded DNA segments by a continuum elastic rod [16, 17]. Longitudinal acoustic vibrations of such a rod fixed at both ends to a static capsid inner surface depend on the longitudinal sound velocity within the rod, which in turn depends on the rod's macroscopic mass density and elastic properties. These two quantities enter the speed of longitudinal sound according to $v = \sqrt{(Y/\rho)}$, where Y is the bulk modulus and ρ is the mass density of the DNA bundle. We can relate these quantities to the microscopic force constant and total genome mass of single double-stranded DNA molecules which have become folded into bundles of length L and circular cross-section area $\pi(L/2)^2$. The elastic response of individual double-stranded DNA molecules has been measured [18]. And elastic constants have been used to calculate phase velocities of sound waves [19]. Brillouin scattering has been used to determine the longitudinal velocity of sound in B-DNA fibers over a quarter century ago [20] with the result v = 1.9 km/s, which we shall adopt for the present calculation. The longitudinal sound velocity in a bundle of folded DNA segments  regarded as a continuum coincides with the sound velocity of a single fiber since Y and ρ have the same ratio for the bundle as for a single constituent  DNA molecule.

We introduce a coordinate system with its origin on the cylindrical bundle's axis where it connects with a static capsid mantle. The x-axis coincides with the bundle's axis and extends through it to the other end of the tightly packaged bundle on the static capsid's internal surface a distance L away. Then with y(x,t) representing a general longitudinal displacement field, for t > 0 in a continuum bundle we have

$$y(x,t) = \sum_n \alpha_n(t) \sin(n\pi x / 2L) \ , \qquad 1.$$

which obeys boundary conditions y(0,t) = 0 and y(L,t) = 0 , where n are even integers and $\alpha_n(t)$ are normal coordinates for longitudinal standing wave motion in the folded DNA bundle, regarded as a continuum.

We take $y(x,t) = 0$ for $t < 0$. The total hamiltonian H of the wave system is given by the sum of its kinetic and potential energy:

$$H = \int_0^L dx \, [s\rho/2] [(\partial y/\partial t)^2 + v^2 (\partial y/\partial x)^2] \, , \qquad 2.$$

where $s = \pi(L/2)^2$ is the cross-sectional area of the packaged DNA bundle, $\rho$ is the mass density and $v$ is the longitudinal sound velocity.

We substitute Eq.(1) into Eq.(2) and obtain

$$H = (s\rho L/4) \sum_n |\dot{\alpha}_n|^2 + (v^2 \rho s \pi^2/16L) \sum_n |\alpha_n|^2 n^2 \, . \qquad 3.$$

The total energy H depends only quadratically on the $\alpha_n(t)$ and their time derivatives. If we now adopt a canonical ensemble to obtain the thermal average $\langle H \rangle$ at temperature T, each quadratic term in H has its equipartition average value $kT/2$ for the ensemble. We obtain:

$$\langle |\alpha_n|^2 \rangle = [8LkT/(n^2 v^2 \rho s \pi^2)] \, , \qquad 4.$$

where the brackets denote the thermal average.

At the closed end pressure antinode, $x = 0$, the pressure fluctuation against the constraining internal wall of the capsid is

$$\Delta p(0,t) = -v^2 \rho \, (\partial y/\partial x \, |_{x=0}) \qquad 5.$$

for a displacement field $y(x,t)$. We now take the ensemble average $\langle |\Delta p|^2 \rangle$ and obtain

$$\langle |\Delta p|^2 \rangle = [v^4 \rho^2 \pi^2/(4L^2)] \sum_{nm} nm \langle \alpha_n \alpha_m \rangle \, . \qquad 6.$$

But $\langle \alpha_n \alpha_m \rangle = \langle |\alpha_n|^2 \rangle \delta_{nm}$ since in thermal equilibrium the normal coordinates $\alpha_n$ are uncorrelated with respect to their time dependence. A single sum results:

$$\langle |\Delta p|^2 \rangle = [v^4 \rho^2 \pi^2/(4L^2)] \sum_n n^2 \langle |\alpha_n|^2 \rangle \, . \qquad 7.$$

Substituting Eq.(4) into Eq.(7) we obtain

$$\langle |\Delta p|^2 \rangle = [2v^2 \rho kT/(Ls)] \sum_n 1 \, . \qquad 8.$$

There must be cut-off limits in Eq.(8) reflecting the requirement that the continuum standing waves entering Eq.(1) have wavelengths that neither exceed the bending persistence length of DNA, 50 nm [21,22,23], and therefore have coherence over the segment length L, nor are shorter than several base pair separations which characterize discrete molecular structure.

For any even integer n the corresponding standing wavelength is given by $\lambda = 4L/n$. Thus we require an even integer N that leads to a wavelength $\lambda_N = 4L/N$ that is no smaller than twice the bundle length L in order to have a lowest coordinate value node at x=L, which is consistent with a bundle segment length L that is equal or less than the bend persistence length. From $4L/N \geq 2L$ we have N=2 . From L ≤ bend persistence length we require L≤ 50 nm. Cited values of total capsid volume, $V_{capsid}$ [6], and corresponding fractions of capsid volume packaged with genomes for three mutants of lambda phage [6] enable us to calculate the respective genome volumes $V_{genome} = V_{\lambda cI60}$, $V_{EMBL3}$, and $V_{\lambda b221}$. And from these we obtain for our surrogate cylindrical capsid model the genome volumes (Ls) and finally L = 40.38 nm, 38.34 nm and 37.13 nm for folded segment lengths of the three respective lambda phage mutants $\lambda c160$, EMBL3, and $\lambda b221$. These values of segment length are all less than the DNA persistence length.

In the Eq.(8) summation we then have N=2, and

$$\sum_n^N 1 = N/2 = 1 \qquad 9.$$

with the thermal fluctuation noise pressure on the capsid mantle then given by:

$$\langle |\Delta p|^2 \rangle = [2v^2 \rho\, kT / (Ls)] . \qquad 10.$$

Taking M=ρ(Ls) as the total genome mass within the capsid, we define the root-mean-square fluctuating pressure magnitudes $P_{rms} \equiv \sqrt{(\langle |\Delta p|^2 \rangle)}$ and obtain

$$P_{rms} \equiv \sqrt{[2v^2 MkT/(V_{genome})^2]} , \qquad 11.$$

where $V_{genome} \equiv (Ls)$ now represents the genome volume for the respective mutant .

Since M is proportional to the genome length, Eq.(11) suggests that the fluctuating pressure magnitude at temperature T should be proportional to the square root of the genome length and inversely proportional to the volume it occupies within the capsid.
.
It has been observed that exposure to increasing osmotic pressure difference between the inside and outside of a lambda bacteriophage capsid can suppress the ejection of the viral genome into a bacterial cytoplasm with an osmotic pressure of several atmospheres [24,25]. These measurements in vitro have shown that a sufficiently large osmotic pressure from outside of a lambda phage caspid can provide a resisting force that balances the internal ejection force exerted by a tightly fitting DNA genome. Relevant data given for three lambda phage mutants [6] using a parameter-free model includes their respective base pair lengths.in addition to total capsid volume and respective fractions of capsid volume occupied by mutant genomes.The osmotic pressures required to completely inhibit the ejection of mutant viral genomes when the capsid is appropriately stimulated are presented [6] as well as lesser osmotic pressures needed to partially prevent DNA ejection. In particular, the published figures indicate complete blockage of 48.5 kbp $\lambda cI60$ ejection at 30 atm, complete blockage of 41.5 kbp EMBL3 ejection at 17 atm, and complete blockage of 37.7 kbp $\lambda b221$ ejection at 12.5 atm osmotic counterpressures.

Following [6] we shall assume these counterpressures match those arising internally within the capsid, at least in part from tightly packaged genomes.

We have calculated the genome volumes $V_{genome}$ from the cited data [6], and the masses M of the three mutants λcI60, EMBL3, and λb221 have been related to their base pair lengths using an average base pair mass of $1.021 \times 10^{-21}$ grams. Using a velocity v = 1.9 km/s [20] to evaluate $P_{rms}$ in Eq.(11) we obtain for T = 310 K:

$$P_{rms,\lambda cI60} = 7.468 \text{ atm},$$

$$P_{rms,EMBL3} = 8.062 \text{ atm},$$

$$P_{rms,\lambda b221} = 8.467 \text{ atm}.$$

The thermal fluctuation pressure component acquires greater values for the less voluminous genome, in spite of its lesser mass and length in base pairs. However, the dependence of total ejection forces on genome length has also been noted in bacteriophage lambda [6], and is indeed greater for longer lengths.

In consequence of the above calculations we conjecture that time-dependent fluctuation pressures leading to $P_{rms}$ act like a trigger that assists the internal forces exerted on tightly packaged DNA to eject the genome in a sequence of steps that follow stimulation by encounter with a receptor on a host cell.

I wish to thank Professor Lisa Lapidus for several stimulating conversations.

.